\documentclass[
aps,
prb,
showpacs,
floatfix,
reprint,
twoside,
superscriptaddress
]{revtex4-1}
\usepackage{amsmath}
\usepackage[T1]{fontenc}
\usepackage{fourier}
\usepackage{graphicx}
\usepackage[colorlinks=true,
allcolors=blue,
dvipdfm=true]{hyperref}

\begin{document}

\title{Excess current in ferromagnet-superconductor structures with fully polarized triplet component}

\author{Andreas Moor}
\affiliation{Theoretische Physik III, Ruhr-Universit\"{a}t Bochum, D-44780 Bochum, Germany}
\author{Anatoly F.~Volkov}
\affiliation{Theoretische Physik III, Ruhr-Universit\"{a}t Bochum, D-44780 Bochum, Germany}
\author{Konstantin B.~Efetov}
\affiliation{Theoretische Physik III, Ruhr-Universit\"{a}t Bochum, D-44780 Bochum, Germany}
\affiliation{National University of Science and Technology ``MISiS'', Moscow, 119049, Russia}

\begin{abstract}
We study the $I$\nobreakdash-$V$~characteristics of S$_{\text{T}}$/n/N~contacts, where S$_{\text{T}}$~is a BCS superconductor~S with a built-in exchange field~$h$, n~represents a normal metal wire, and N---a normal metal reservoir. The superconductor S$_{\text{T}}$ is separated from the n\nobreakdash-wire by a spin filter which allows the passage of electrons with a certain spin direction so that only fully polarized triplet Cooper pairs penetrate into the n\nobreakdash-wire. We show that both the subgap conductance~$\sigma_{\text{sg}}$ and the excess current~$I_{\text{exc}}$, which occur in conventional S/n/N~contacts due to Andreev reflection~(AR), exist also in the considered system. In our case, they are caused by unconventional~AR that is not accompanied by spin flip. The excess current~$I_{\text{exc}}$ exists only if~$h$ exceeds a certain magnitude~$h_{\text{c}}$. At ${h < h_{\text{c}}}$ the excess current is converted into a deficit current~$I_\text{{def}}$. The dependencies of the differential conductance and the current~$I_{\text{exc}}$ are presented as a function of voltage and~$h$.
\end{abstract}

\date{\today}

\maketitle

\section{Introduction}

As is well known, a so-called excess current~$I_{\text{exc}}$ appears at large voltages~$V$ in Josephson junctions (JJ) with a direct conductance,\cite{Likharev,Barone_Paterno_1982} that is, the current~$I_{\text{exc}}$ arises in JJs of the S/n/S or S/c/S types, where~n denotes a normal metal (a wire or a film) and~c---a constriction. This means that the current-voltage ($I$\nobreakdash-$V$) characteristics at large~$V$ (${e V \gg \Delta}$, where $\Delta$ is the energy gap in the superconductors~S) has the form
\begin{equation}
    I(V) = V / R + I_{\text{exc}} \mathrm{sgn}(V) \,, \label{1}
\end{equation}
where~$R$ is the resistance of the JJ in the normal state and the constant~$I_{\text{exc}}$ is the excess current which can be written in the form
\begin{equation}
    I_{\text{exc}} = a \Delta / R \,. \label{1'}
\end{equation}
Here, $a$~is a numerical factor equal to ${a = \pi^{2} / 4 - 1}$ in the diffusive limit,\cite{AVZ79} and ${a = 8/3}$ in ballistic JJs with ideal (fully transparent) interfaces.\cite{Zaitsev80,*Zaitsev80a,BTK82} Eq.~(\ref{1}) also describes the asymptotic behavior (${eV \gg \Delta}$) of the $I$\nobreakdash-$V$ characteristics of S/n/N contacts,\cite{AVZssc79,Zaitsev80,BTK82} where~N is a normal metal reservoir. In the latter case, the excess current is twice smaller than in the S/n/S~JJs. The excess current~$I_{\text{exc}}$ is an essential characteristics of~S/n/N or~S/n/S contacts which distinguishes them from the tunnel junctions~S/I/N or~S/I/S where this current does not arise.

If the S/n or n/N~interfaces are not ideal (the transmission coefficient differs from~1), the coefficient~$a$ in Eq.~(\ref{1'}) can be either positive or negative. That is, an excess~$I_{\text{exc}}$ or deficit~$I_{\text{def}}$ currents arise in this case. Their values depend on the interface transparencies of both interfaces.\cite{VOLKOV199321} The appearance of the excess current at large~$V$ as well as the non-zero subgap conductance~$G(V,T)$ of the S/n/N~contacts at ${V \leq \Delta / e}$ and ${T = 0}$ is explained\cite{AVZssc79,Zaitsev80,*Zaitsev80a,BTK82} in terms of Andreev reflections~(AR).\cite{Andreev64} It has been shown in Refs.~\onlinecite{AVZssc79,Zaitsev80,*Zaitsev80a,BTK82} that the zero bias conductance~$G(0,0)$ coincides with the conductance in the normal state and has a non-monotonous dependence on the applied voltage~$V$ or temperature~$T$. Similar behavior of the conductance takes place in the so-called Andreev interferometers (see experimental observations in Refs.~\onlinecite{Pothier_et_al_1994,Vegvar_et_al_1994,Dimoulas_et_al_1995,Petrashov95} and theoretical explanations in Refs.~\onlinecite{Volkov_et_al_1996_L45,Nazarov_Stoof_1996}).

The Andreev reflection implies that an electron moving in the normal metal towards the superconductor is converted at the S/n~interface into a hole with opposite spin which moves back along the same trajectory. Physically, this process means that an electron with momentum~$\mathbf{p}$ and spin~$\mathbf{s}$ moving from the n\nobreakdash-metal penetrates the superconductor~S and forms there a Cooper pair, i.e., it pulls another electron with opposite momentum~$-\mathbf{p}$ and spin~$-\mathbf{s}$. The absence of this electron in the n\nobreakdash-metal is nothing else as the creation of a hole with momentum~$-\mathbf{p}$ and spin~$-\mathbf{s}$. In the superconductor/ferromagnet (S/F)~contacts, the AR is suppressed since the exchange field~$h$ acting on spins breaks the symmetry of spin directions. De Jong and Beenakker\cite{de_Jong_Beenakker_1995} have shown that the conductance~$G(V,T)_{|_{V=T=0}}$ in ballistic S/F~systems is reduced with increasing~$h$ and turns to zero at ${h > E_{\text{F}}}$, where~$E_{\text{F}}$ is the Fermi energy. At high exchange energy, electrons with only one spin direction exist in the ferromagnet~F so that the AR at S/F interfaces is not possible.

One can expect a similar behavior of the conductance in S$_{\text{T}}$/n/N~contacts, where a ``magnetic'' superconductor with a spin filter S$_{\text{T}}$ (see below) supplies only fully polarized triplet Cooper pairs penetrating the n\nobreakdash-metal. It consists of an S/F~bilayer and a spin filter~Fl which passes electrons with only one spin direction, so that one deals with the S$_{\text{T}}$ superconductor constructed as a multylayer structure of the type S/F/Fl. In this case, the conventional AR at the S$_{\text{T}}$/n~interface is forbidden and, therefore, the subgap conductance at low temperatures as well as the excess current may disappear.

As will be shown in this work, the subgap conductance as well as the excess current~$I_{\text{exc}}$ remain finite in S$_{\text{T}}$/n/N~contacts. The magnitude of the current~$I_{\text{exc}}$ and its sign depend on the value of the exchange field in the ferromagnet~F. In the considered case of S$_{\text{T}}$/n/N~contacts, the subgap conductance and the excess current occur due to an unconventional AR in which two electrons with parallel spins in the n\nobreakdash-film form a triplet Cooper pair with the same direction of the total spin. Therefore, the AR at the S$_{\text{T}}$/n interface is not accompanied by spin-flip (the hole in the n\nobreakdash-wire has the same spin direction as the incident electron).

Note that, nowadays, the interest in studies of the excess current is revived in the light of recent measurements on S/Sm/S~JJs with unconventional semiconductor~Sm (topological insulator) in which the Josephson effect can occur due to Majorana modes (see recent experimental papers Refs.~\onlinecite{Marcus15,Klapwijk16}, and references therein). In these junctions, the excess current also has been observed. On the other hand, properties of high-$T_{\text{c}}$ superconductors including the iron-based pnictides have been also studied with the aid of point-contact spectroscopy in which the differential conductance of N/S~point contacts has been measured.\cite{Chen08,Kuroki10,Noat10,Sheet10,Shan11} A theory of differential conductance of N/S~point contacts composed by a two band superconductor with energy gaps of different signs [${\mathrm{sgn}(\Delta_{1}) = -\mathrm{sgn}(\Delta_{2})}$] has been presented in Ref.~\onlinecite{Vavilov11}.

In this Paper, we calculate the $I$\nobreakdash-$V$~characteristics of diffusive superconductor/normal metal systems of two types. In the first type of contacts, S$_{\text{m}}$/n/N, the ``magnetic'' superconductor S$_{\text{m}}$ is a singlet superconductor~S covered by a thin ferromagnetic layer [see Fig.~\ref{fig:System1a}~(a)]. In this case, both the singlet and the triplet Cooper pairs penetrate into the n\nobreakdash-wire. In the second type of contacts, S$_{\text{T}}$/n/N, the magnetic superconductor~S$_{\text{T}}$ consists again of an S/F~bilayer which is separated from the n\nobreakdash-wire by a spin filter~Fl [see Fig.~\ref{fig:System1a}~(b)]. The spin filter~Fl is assumed to pass only electrons with spins oriented along the $z$~axis (${\mathbf{s} || \hat{\mathbf{z}}}$). Using the quasiclassical theory, we show that in both types of contacts, S$_{\text{m}}$/n/N and S$_{\text{T}}$/n/N, the conductance~$G$ is affected by the proximity effect and the excess (deficit) current~$I_{\text{exc}}$ ($I_{\text{def}}$) as well as the subgap conductance are finite.

\section{Model and Basic Equations}

\begin{figure}[tbp]
\includegraphics[width=1.0\columnwidth]{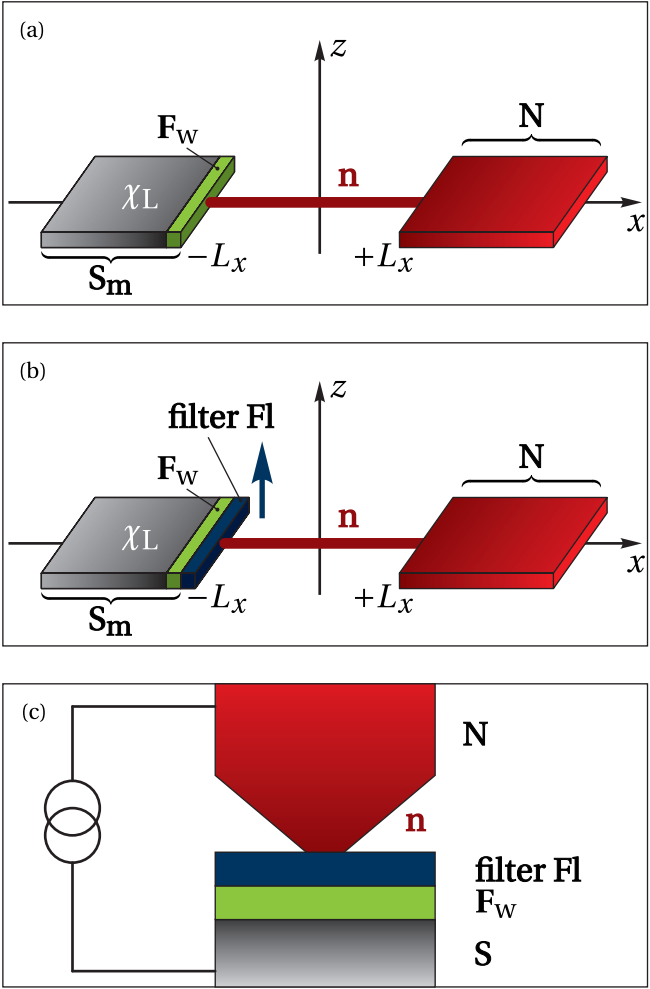}
\caption{(Color online.) Schematic representation of the system under consideration (not to scale). (a)~S$_{\text{m}}$/n/N contact---the superconductor S$_{\text{m}}$ consists of a BCS superconductor~S and a thin ferromagnetic layer (denoted by F$_{\text{w}}$), and is connected to a normal metal reservoir~N on the right hand side via a normal metal wire~n. (b)~S$_{\text{T}}$/n/N contact---in addition to the case~(a), the S$_{\text{m}}$ superconductor on the left hand side is covered by a spin filter~Fl that passes electrons only with a certain spin direction, say, parallel or antiparallel to the $z$~axis (indicated by the thick blue arrow). The superconducting phase on the left hand side is~$\chi_{\text{L}}$. (c)~Sketch (not to scale) of a possible experimental realization of the case~(b).}
\label{fig:System1a}
\end{figure}

We consider an S$_{\text{T}}$/n/N contact, in which the ``magnetic'' superconductors are formed by a BCS superconductor~S (s\nobreakdash-wave, singlet) covered by a thin ferromagnetic layer~F with an exchange field~$\mathbf{h}$ [see Fig.~\ref{fig:System1a}~(a)]. Due to proximity effect, the singlet component penetrates from the superconductor into the F~film, and also a triplet component arises under the action of the exchange field~$\mathbf{h}$. As is well known (see reviews Refs.~\onlinecite{BuzdinRMP,BVErmp,Eschrig_Ph_Today,Eschrig_Reports_2015}), in the case of homogeneous magnetization~$\mathbf{M}$ (${\mathbf{M} || \mathbf{h}}$) in the ferromagnet, the vector of the total spin of triplet Cooper pairs~$\mathbf{S}$ lies in the plane perpendicular to~$\mathbf{M}$. Thus, the S/F~bilayer with a sufficiently transparent interface can be considered as a ``magnetic'' superconductor with a built-in effective exchange field~$\mathbf{h}_{\text{eff}}$ which has a nonzero projection onto the $z$~axis and an effective energy gap~$\Delta_{\text{eff}}$ (to be more exact, the condensate wave functions in the F~film are analogous to those in a ``magnetic'' superconductor).

The magnitudes of ${h_{\text{eff}} = \mathbf{h}_{\text{eff}}}$ and $\Delta_{\text{eff}}$ are determined by certain conditions. For example, in case of thin~F and~S layers (${d_{\text{F}} \ll \xi_{h}}$, ${d_{\text{S}} \ll \xi_{S}}$, where~$d_{\text{F,S}}$ are the thicknesses of the F(S)~layers, ${\xi_{h} = \sqrt{D/h}}$ and ${\xi_{S} = \sqrt{D/\Delta}}$) and a low F/S~interface resistance, one has\cite{Bergeret_Volkov_Efetov_2001_b}
\begin{align}
h_{\text{eff}}       &= h \frac{\nu_{\text{F}} d_{\text{F}}}{\nu_{\text{F}} d_{\text{F}} + \nu_{\text{S}} d_{\text{S}}} \,, \label{1a} \\
\Delta_{\text{eff}} &= \Delta_{\text{S}} \frac{\nu_{\text{S}} d_{\text{S}}}{\nu_{\text{F}} d_{\text{F}} + \nu_{\text{S}} d_{\text{S}}} \,.
\end{align}

In case of a high S/F interface resistance, we obtain (see the Appendix~\ref{app:Appendix_A})
\begin{align}
h_{\text{eff}} &= h \,, \\
\Delta_{\text{eff}} &=\epsilon_{\text{sg}} \equiv \frac{D \kappa_{\text{SF}}}{d_{\text{F}}} \,, \label{1b}
\end{align}
where~$D$ is the diffusion coefficient in the F~film, ${\kappa_{\text{SF}} = (\sigma R_{\text{SF}})^{-1}}$, $\sigma$~is the conductivity of the F~film and~$R_{\text{SF}}$ is the S/F interface resistance per unit area [see below Eqs.~(\ref{4}) and~(\ref{4'})]. The quantity~$\epsilon_{\text{sg}}$ determined by the interface resistance is the so-called subgap or minigap.\cite{McMillan_1968} In both cases, the effective exchange field~$h_{\text{eff}}$ may exceed the effective gap~${h_{\text{eff}}}$ without causing a non-uniform state of the Larkin--Ovchinnikov--Fulde--Ferrel type\cite{Larkin_Ovchinnikov_1965,Fulde_Ferrell_1964} because the thicker S~film is only weakly affected by the F~film. For example in the famous experiment,\cite{Birge10} where a long-range triplet component has been observed in a multilayered S/F'/F/F'/S Josephson junction, the Curie temperature in a weak ferromagnet~F' (Pd$_{0.88}$Ni$_{0.12}$) was about~$175~\text{K}$, that is, much larger than the critical temperature of the superconducting transition in the superconductor~S (Nb) with a transition temperature ${T_{\text{c}} = 9~\text{K}}$. In principle, a similar F'/S bilayer can be employed as a prototype of the presented S$_{\text{m}}$~superconductor.

The F~layer in S$_{\text{T}}$/n/N~contacts is separated from the n\nobreakdash-wire (or film) by a filter that passes electrons only with a certain spin direction, say, parallel or antiparallel to the $z$~axis [see Fig.~\ref{fig:System1a}~(b)]. As a filter, thin layers of strongly polarized magnetic insulator\cite{Moodera11,*Moodera_1993,Fert06,Ramos07} and DyN or GdN films\cite{Muduli_et_al_arXiv_2015} can be used.

The convenient method to study the system under consideration is the theory of quasiclassical Green's functions.\cite{RammerSmith,LO,BelzigRev,Kopnin} This technique is generalized for the case of ferromagnet-superconductor structures where a non-trivial dependence of the quasiclassical Green's functions~$\check{g}$ on spin indices must be taken into account.\cite{BuzdinRMP,BVErmp,Eschrig_Reports_2015} In the considered non-equilibrium case, the Green's function~$\check{g}$ is a matrix with diagonal matrix elements ($\hat{g}^{R}$ and $\hat{g}^{A}$) and non-diagonal element~($\hat{g}^{K}$), where the matrices~$\hat{g}^{R(A)}$ and $\hat{g}^{K}$ are the retarded (advanced) and Keldysh functions, respectively. All these functions are ${4 \times 4}$ matrices in the Gor'kov-Nambu and spin spaces.

In the n\nobreakdash-wire the matrix~$\check{g}$ obeys an equation which looks similar to the Usadel equation\cite{Usadel} (see also Eq.~(5) in Ref.~\onlinecite{VOLKOV199321})
\begin{equation}
    \nabla (\check{g} \nabla \check{g}) + i \kappa_{\epsilon}^{2} [\hat{X}_{30} \,, \check{g}] = 0 \,, \label{2}
\end{equation}
where ${\kappa_{\epsilon}^{2} = \epsilon / D}$ with the diffusion coefficient~$D$. The matrix~${\hat{X}_{30} = \hat{\tau}_{3} \cdot \hat{\sigma}_{0}}$ is a tensor product of the Pauli matrices $\hat{\tau}_{i}$ (${i = 1,2,3}$) and the ${2 \times 2}$ unit matrix~$\hat{\sigma}_{0}$, which operate in the particle-hole and spin space, respectively. The matrix quasiclassical Green's function~$\check{g}$ obeys the normalization condition
\begin{equation}
    \check{g} \cdot \check{g} = \check{1} \,. \label{3}
\end{equation}
Equation~(\ref{2}) is complemented by boundary conditions at the interfaces S$_{\text{m}}$/n and n/N. They have the form [see Eq.~(4.7) in Ref.~\onlinecite{Bergeret12a,*Bergeret12b}, Refs.~\onlinecite{EschrigBC13,EschrigBC13a,EschrigBC15}, and also the recent work~Ref.~\onlinecite{Machon_Belzig_2015}]
\begin{align}
2 \bar{r}_{\text{S}} L \check{g} \partial_{x} \check{g} &= [\hat{\Gamma} \hat{G} \hat{\Gamma} \,, \check{g}]|_{0} \,, \label{4} \\
2 \bar{r}_{\text{N}} L \check{g} \partial_{x} \check{g} &= [\check{g}\,, \hat{G}_{\text{N}}]|_{L} \,. \label{4'}
\end{align}
Here, the sub-indices~$0$ and~$L$ relate to the n/S$_{\text{m}}$ and n/N~interfaces, respectively, while ${\bar{r}_{\text{S},\text{N}} = \sigma R_{\text{S},\text{N}} / L}$, where~$\sigma$ is the conductivity of the n\nobreakdash-wire, and~$R_{\text{S},\text{N}}$ denote the S$_{\text{m}}$/n (respectively, n/N) interface resistance per unit area. The matrix~$\hat{\Gamma}$ describes the electron transmission with a spin-dependent probability~$\mathcal{T}_{\uparrow,\downarrow}$. If the filters let to pass only electrons with spins parallel to the $z$~axis, then ${\hat{\Gamma} = \mathcal{T} \hat{1} + \mathcal{U} \hat{X}_{33}}$ so that the probability for an electron with spin up (down) to pass into the n\nobreakdash-wire is ${\mathcal{T}_{\uparrow,\downarrow} \propto \mathcal{T} \pm \mathcal{U}}$. We assume that ${\mathcal{U} = \zeta \mathcal{T}}$ with ${\zeta = \pm 1}$, and the coefficients~$\mathcal{T}$ and $\mathcal{U}$ are normalized, i.e., ${\mathcal{T} = |\mathcal{U}| = \sqrt{2}}$. Note that coefficients~$\bar{r}_{\text{S},\text{N}}$ are inverted with respect to the coefficients~$r_{\nu}$ used in Refs.~\onlinecite{Moor_Volkov_Efetov_2015_c,Moor_Volkov_Efetov_2015_d}.

Consider first Eq.~(\ref{2}) for the Keldysh Green's function $\hat{g}^{K}$. In the considered one-dimensional case it has the form
\begin{equation}
    \partial_{x}(\hat{g}^{R} \partial_{x} \hat{g}^{K} + \hat{g}^{K} \partial_{x} \hat{g}^{A}) + i \kappa_{\epsilon}^{2} [\hat{X}_{30} \,, \hat{g}^{K}] = 0 \,. \label{5K}
\end{equation}
The Keldysh function~$\hat{g}^{K}$ can be expressed in terms of the retarded and advanced Green's functions $\hat{g}^{R(A)}$, and the matrix distribution function ${\hat{n} = n_{l} \hat{X}_{00} + n \hat{X}_{30}}$,
\begin{equation}
    \hat{g}^{K} = \hat{g}^{R} \cdot \hat{n} - \hat{n} \cdot \hat{g}^{A} \,. \label{5f}
\end{equation}
The distribution function~$n_{l}$ determines the superconducting order parameter~$\Delta$, whereas the function~$n$ describes the dissipative current.\cite{Schmid75,ArtVolkovRev80} We need to know only the distribution function~$n$. Multiplying Eq.~(\ref{5K}) by~$\hat{X}_{30}$ and taking trace we obtain (employing the normalization condition Eq.~(\ref{3}), in particular, the relations ${\hat{g}^{R(A)} \cdot \hat{g}^{R(A)} = \hat{1}}$)
\begin{equation}
    \big[ 1 - (\hat{g}_{||}^{R} \cdot \hat{g}_{||}^{A})_{00} + (\hat{g}_{\perp}^{R} \cdot \hat{g}_{\perp}^{A})_{00} \big] \partial_{x} n = J \,, \label{6f}
\end{equation}%
where~$\hat{g}_{||, \perp}^{R(A)}$ are, respectively, the diagonal and off-diagonal elements of~$\hat{g}^{R(A)}$ matrices in the particle-hole space, and we introduced the notation ${(\ldots)_{ij} = \mathrm{Tr} \{ \hat{X}_{ij} (\ldots) \}/4}$. The quantity ${J = J(\epsilon)}$ is independent of~$x$. Integrating Eq.~(\ref{6f}) we obtain
\begin{equation}
    n(x) = n_{0} + J \int_{0}^{x} \frac{\mathrm{d} x}{1 + M_{n}(x)} \,, \label{7f}
\end{equation}
where ${M_{n}(x) = - (\hat{g}_{||}^{R} \cdot \hat{g}_{||}^{A})_{00} + (\hat{g}_{\perp}^{R} \cdot \hat{g}_{\perp}^{A})_{00}}$.

Using the boundary conditions, Eqs.~(\ref{4}) and~(\ref{4'}), we find
\begin{equation}
    J L = \frac{F_{V}}{2 \bar{r}_{\text{N}} / M_{\text{N}} + 2 \bar{r}_{\text{S}} / M_{\text{S}} + \big\langle( 1 + M_{n}(x) )^{-1} \big\rangle} \,, \label{8J}
\end{equation}%
where ${F_{V} = (1/2) \big[\tanh [(\epsilon + eV)/2T] - \tanh [(\epsilon - eV)/2T] \big]}$ is the distribution function in the normal metal reservoir (we set the voltage in the S~reservoir equal to zero), ${\langle \ldots \rangle \equiv L^{-1} \int_{0}^{L}(\ldots)}$, ${M_{\text{S}} = \big((\hat{g}^{R} - \hat{g}^{A})_{||} (\hat{G}_{\text{S}}^{R} - \hat{G}_{\text{S}}^{A})_{||} + (\hat{g}^{R} + \hat{g}^{A})_{\perp} (\hat{G}_{\text{S}}^{R} + \hat{G}_{\text{S}}^{A})_{\perp }\big)_{00}}$, and ${M_{\text{N}} = \big( (\hat{g}^{R} - \hat{g}^{A})_{||} (\hat{G}_{\text{N}}^{R} - \hat{G}_{\text{N}}^{A})_{||} \big)_{00}}$.

The current~$I$ is expressed via the ``partial'' current~$J$ as
\begin{equation}
    I = (\sigma/4eL) \int J(\epsilon) \, \mathrm{d} \epsilon \,. \label{9I}
\end{equation}

Formula Eq.~(\ref{8J}) generalizes Eq.~(13) of Ref.~\onlinecite{VOLKOV199321} for the considered case of a spin-dependent interaction and can be applied to the description of contacts with a condensate consisting of singlet and triplet Cooper pairs. In the normal state, above the critical temperature of the superconductor~S, one has ${M_{\text{N}} = M_{\text{S}} = 2 (1 + M_{\text{n}}) = 4}$. Thus, we obtain a standard expression for the current per unit area in an N/n/N~contact
\begin{equation}
    I = \frac{V}{R_{\text{S}} + R_{\text{N}} + L / \sigma} \,. \label{10J}
\end{equation}
The denominator is the sum of interface resistances and the resistance of the normal n\nobreakdash-wire.

The normalized differential conductance of the contacts under consideration ${\tilde{\sigma}_{\text{d}}(v) \equiv (\mathrm{d}I/\mathrm{d}V)/\sigma_{\text{N}}}$ at ${T = 0}$ is
\begin{equation}
    \tilde{\sigma}_{\text{d}}(v) = \frac{(\bar{r}_{\text{S}} + \bar{r}_{\text{N}} + 1) / 4}{\bar{r}_{\text{N}} / M_{L}(eV) + \bar{r}_{\text{S}} / M_{0}(eV) + \big\langle( 1 + M_{n}(x,eV))^{-1} \big\rangle / 2} \,, \label{11dif}
\end{equation}
where ${v = eV/\Delta}$ is the normalized voltage. The normalized current ${\tilde{I}(v) \equiv I(eV/\Delta)(L/\sigma V)}$ is given by the relation ${\tilde{I}(v) = \int_{0}^{v}\tilde{\sigma}_{\text{d}}(v_{1}) \mathrm{d}v_{1}}$ and, at large voltage, can be written in the form
\begin{equation}
    \tilde{I}(v) = \tilde{I}_{\text{N}}(v) + \delta \tilde{I}(v) \,, \label{12exc}
\end{equation}
where ${\tilde{I}_{\text{N}}(v) = v / (1 + \bar{r}_{\text{S}} + \bar{r}_{\text{N}})}$ is the normalized current through the contact in the normal state. The normalized excess (${\delta \tilde{I} = \tilde{I}_{\text{exc}}}$) or deficit current (${\delta \tilde{I} = \tilde{I}_{\text{def}}}$) is determined by the expression
\begin{equation}
    \delta \tilde{I} \equiv \delta \tilde{I}(\infty) = \int_{0}^{\infty} [\tilde{\sigma}_{\text{d}}(v) - 1] \, \mathrm{d} v \,. \label{13exc}
\end{equation}
It is valid at arbitrary temperatures because for the function~$F_V$ in Eq.~(\ref{8J}) we have ${F_V \to 1}$ for ${V \to \infty}$.

The current~$\tilde{I}(v)$ can be presented as ${\tilde{I}(v) = \tilde{I}_{<} + \tilde{I}_{>}}$, where ${\tilde{I}_{<} = \int_{0}^{1} \sigma_{\text{d}}(v_{1}) \, \mathrm{d} v_{1} }$ is a subgap current and ${\tilde{I}_{>} = \int_{1}^{v} \sigma_{\text{d}}(v_{1}) \, \mathrm{d} v_{1}}$ is the contribution from quasiparticles with energies above the gap; the normalized current in the normal state is ${\tilde{I}_{\text{N}}(1) = (1 + \bar{r}_{\text{S}} + \bar{r}_{\text{N}})^{-1}}$.

We see that the excess current is determined by the retarded (advanced) Green's functions $\hat{g}^{R(A)}$ that obey an Usadel-like equation. This equation can be solved in limiting cases. We consider a contact with a short n\nobreakdash-wire (${L \ll \xi_{\text{S}} \simeq \sqrt{D/\pi T_{\text{c}}}}$) in which the interface resistances dominate (${\bar{r}_{\text{S},\text{N}} \gg 1}$), i.e., the interface resistances are much larger than the resistance of the n\nobreakdash-wire, ${R_{\text{S},\text{N}} \gg L/\sigma}$.

\subsection{Retarded (advanced) Green's Functions}
\label{sec:2a}

In the case of a short contact, the last term in the denominator of Eq.~(\ref{11dif}) and the second term in Eq.~(\ref{5K}) can be neglected so that the Usadel equation for the Green's functions~$\hat{g}^{R(A)}$ acquires the form
\begin{equation}
    \partial_{x}(\hat{g}^{R} \partial_{x} \hat{g}^{R}) = 0 \,, \label{14Usadel}
\end{equation}
provided that ${L \ll \xi_{\text{S}} \simeq \sqrt{D/T_{\text{c}}}}$. We integrate Eq.~(\ref{14Usadel}) once over~$x$ and obtain
\begin{equation}
    \hat{J}^{R(A)} = (\hat{g} \partial_{x} \hat{g})^{R(A)} \,. \label{14a}
\end{equation}
From the boundary conditions Eqs.~(\ref{4}) and~(\ref{4'}) for the retarded (advanced) Green's functions, we have
\begin{align}
    2 \hat{J}^{R(A)} L &= \bar{r}_{\text{S}}^{-1} [ \hat{\Gamma} \hat{G}_{\text{S}} \hat{\Gamma} \,, \hat{g}(0) ]^{R(A)} \,, \label{15J} \\
    2 \hat{J}^{R(A)}L &= - \bar{r}_{\text{N}}^{-1} [ \hat{G}_{\text{N}} \,, \hat{g}(L)]^{R(A)} \,. \label{15J'}
\end{align}
Subtracting the first equation from the second we arrive at
\begin{equation}
    [ \hat{\Lambda} \,, \hat{g} ]^{R(A)} = 0 \,, \label{15Usadel}
\end{equation}
where the matrix ${\hat{\Lambda} = \hat{\Lambda}_{\text{N}} + \hat{\Lambda}_{\text{S}}}$ is a sum of contributions of the n/N and S$_{\text{m}}$/n interfaces, ${\hat{\Lambda}_{\text{N}} = \bar{r}_{\text{N}}^{-1} \hat{X}_{30}}$ and ${\hat{\Lambda}_{\text{S}} = \bar{r}_{\text{S}}^{-1} [ G_{||} \hat{X}_{||} + G_{\perp} \hat{X}_{\perp} ]}$. The form of matrices~$\hat{X}_{||}$ and~$\hat{X}_{\perp}$ depends on the type of a superconductor.

\paragraph{``Magnetic'' superconductor~S$_{\text{m}}$.}

That is, the superconductor~S$_{m}$ is represented by an S/F~bilayer with a thin ferromagnetic layer~F. We assume that the exchange field~$\mathbf{h}$ is aligned parallel to the z~axis, ${\mathbf{h} || \hat{\mathbf{z}}}$. In this case,
\begin{equation}
\hat{\Lambda}_{\text{S}}^{(a)} \equiv \hat{\Lambda}_{\text{\text{S}}_{\text{m}}} = \bar{r}_{\text{S}}^{-1} [ G_{\text{S}+} \hat{X}_{30} + G_{\text{S}-} \hat{X}_{33} + (F_{\text{S}+} \hat{X}_{10} + F_{\text{S}-} \hat{X}_{13}) ] \,, \label{eq:Lambda_a}
\end{equation}
with\cite{Moor_Volkov_Efetov_2015_c,Moor_Volkov_Efetov_2015_d}
\begin{align}
G_{\text{S} \pm}^{R(A)} &= \frac{ [\zeta^{R(A)}(\epsilon + h)]^{-1} |\epsilon + h| \pm [\zeta^{R(A)}(\epsilon - h)]^{-1} |\epsilon - h| }{2} \,, \label{G_S_mag}\\
F_{\text{S} \pm}^{R(A)} &= \frac{\Delta \big[ [\zeta^{R(A)}(\epsilon + h)]^{-1}  \pm [\zeta^{R(A)}(\epsilon - h)]^{-1} \big]}{2}  \,. \label{F_S_mag}
\end{align}
The terms~$F_{\text{S}+} \hat{X}_{10}$ and~$F_{\text{S}-} \hat{X}_{13}$ in Eq.~(\ref{eq:Lambda_a}) describe the singlet component and, respectively, the short-range triplet component with the total spin of triplet Cooper pairs~$\mathbf{S}$ normal to the $\mathbf{h}$~vector.

Note that the energy gap~$\Delta$ and the exchange field~$h$ in Eqs.~(\ref{G_S_mag}) and~(\ref{F_S_mag}) mean the effective~$\Delta_{\text{eff}}$ and~$h_{\text{eff}}$ defined in Eqs.~(\ref{1a})\nobreakdash--(\ref{1b}). In the following, for brevity,  we drop the subindex ``eff''.

\paragraph{``Triplet'' superconductor S$_{\text{T}}$.}

This case can be realized with the help of an S/F~bilayer with the $\mathbf{h}$~vector aligned, for instance, along the $x$~axis. The S/F~bilayer is assumed to be separated from the n\nobreakdash-wire by a spin filter oriented parallel to the $z$~axis. Then,
\begin{equation}
\hat{\Lambda}_{\text{S}}^{(b)} \equiv \hat{\Lambda}_{\text{S}\text{T}} = \bar{r}_{\text{S}}^{-1} \big[ G_{\text{S}+}(\hat{X}_{30} + \hat{X}_{03}) + F_{\text{S}-} (\hat{X}_{11} - \hat{X}_{22}) \big] \,.
\end{equation}
The last term describes fully polarized triplet Cooper pairs with the $\mathbf{S}$~vector oriented along the $z$~axis.

\paragraph{BCS-superconductor.}

For completeness, we consider also the case of the BCS superconductor which is obtained from the case of a ``magnetic'' superconductor S$_{\text{m}}$ setting ${h = 0}$. Here,
\begin{equation}
\hat{\Lambda}_{\text{S}}^{(c)} \equiv \hat{\Lambda}_{\text{BCS}} = \bar{r}_{\text{S}}^{-1} \big[ G_{\text{S}} \hat{X}_{30} + F_{\text{S}} \hat{X}_{10} \big] \,,
\end{equation}
with
\begin{align}
G_{\text{S}}^{R(A)} &= \epsilon \big[ \zeta^{R(A)} \big]^{-1} \,, \\
F_{\text{S}}^{R(A)} &= \Delta \big[ \zeta^{R(A)} \big]^{-1} \,,
\end{align}
and ${\zeta^{R(A)} = \sqrt{(\epsilon \pm i \Gamma)^{2} - \Delta^{2}}}$.

\subsection{General form of $\hat{g}$ in case of large interface resistance}

In order to make the results more transparent, we assume that the parameter~$\bar{r}_{\text{N}}/\bar{r}_{\text{S}}$ is small and both parameters~$\bar{r}_{\text{N},\text{S}}$ are large (${\bar{r}_{\text{N},\text{S}} \gg 1}$). These conditions correspond to experimental systems and mean that the S/n~interface resistance is much larger than the resistance of the n/N~interface and both interface resistances are larger than the resistance of the short n\nobreakdash-wire. Then, the solution for a small correction ${\delta \hat{g}^{R(A)} = \hat{g}^{R(A)} - \hat{g}_{0}^{R(A)}}$ [where ${\hat{g}_{0}^{R(A)} = \pm \hat{X}_{30}}$ are the quasiclassical retarded (advanced) Green's functions in the separated n\nobreakdash-wire] is
\begin{equation}
    \delta \hat{g}^{R(A)} \equiv \delta \hat{f}^{R(A)} \simeq \frac{\bar{r}_{\text{N}}}{\bar{r}_{\text{S}}} \hat{G}_{\perp}^{R(A)} \,. \label{16gUsadel}
\end{equation}
We see that in the lowest approximation in the parameter~$\bar{r}_{\text{N}}/\bar{r}_{\text{S}}$ only the condensate wave function, off-diagonal in the Gor'kov-Nambu space, is changed due to proximity effect. The correction~$\delta \hat{g}^{R(A)}$ is small if the parameter ${\gamma \equiv \bar{r}_{\text{N}}/\bar{r}_{\text{S}}}$ is small or, in the case of the S$_{\text{T}}$/n/N~contact, if the parameter~$h/\Delta$ is small.

\section{Differential conductance and the $I$-$V$~curve in a short contact}

\subsection{Differential Conductance}

Using the known function ${\hat{g}^{R(A)} = \hat{g}_{0}^{R(A)} + \delta \hat{g}^{R(A)}}$ and Eq.~(\ref{11dif}), we can readily calculate the normalized conductance $\tilde{\sigma}_{\text{d}}(v)$ at ${T = 0}$. Thus, we obtain
\begin{equation}
    \tilde{\sigma}_{\text{d}}(v) = \frac{1 + \gamma}{\gamma + [\nu_{S} + \gamma f^{2}]^{-1}}|_{\epsilon = v} \,, \label{17Con}
\end{equation}
with the functions
\begin{align}
    \nu_{\text{S}} &=
            \begin{cases}
                \frac{ \Re\big\{ [\zeta^{R}(\epsilon + h)]^{-1} |\epsilon + h| + [\zeta^{R}(\epsilon - h)]^{-1} |\epsilon - h| \big\} }{2} \,, & \text{S$_{\text{m}}$/n/N} \,, \\
                \frac{ \Re\big\{ [\zeta^{R}(\epsilon + h)]^{-1} |\epsilon + h| + [\zeta^{R}(\epsilon - h)]^{-1} |\epsilon - h| \big\} }{2} \,, & \text{\text{S$_{\text{T}}$/n/N}} \,, \\
                \Re\Big\{ \frac{|\epsilon|}{\zeta^{R}_{+}(\epsilon)} \Big\} \,, & \text{S/n/N} \,,
            \end{cases}
\end{align}
and
\begin{align}
f^{2} &= \frac{\mathrm{Tr} \big\{ (\hat{G}_{\perp}^{R} + \hat{G}_{\perp}^{A})^{2} \big\}}{16} \\
    &=  \begin{cases}
            \frac{\Big[ \Re \Big\{ \frac{\Delta}{\zeta^{R}(\epsilon + h)} + \frac{\Delta}{\zeta^{R}(\epsilon - h)} \Big\} \Big]^2 + \Big[ \Re \Big\{ \frac{\Delta}{\zeta^{R}(\epsilon + h)} - \frac{\Delta}{\zeta^{R}(\epsilon - h)} \Big\} \Big]^2}{2} \,, & \text{S$_{\text{m}}$/n/N} \,, \\
            \frac{\Big[ \Re \Big\{ \frac{\Delta}{\zeta^{R}(\epsilon + h)} - \frac{\Delta}{\zeta^{R}(\epsilon - h)} \Big\} \Big]^2}{2} \,, & \text{\text{S$_{\text{T}}$/n/N}} \,, \\
            \Big[ \Re\Big\{ \frac{\Delta}{\zeta^{R}_{-}(\epsilon)} \Big\} \Big]^2 \,, & \text{S/n/N} \,,
        \end{cases} \notag
\end{align}
where ${\zeta^{R}_{\pm}(\epsilon) = \sqrt{\pm [(\epsilon + i \Gamma)^2 - \Delta^2]}}$. Equation~(\ref{17Con}) determines the dependence of the normalized differential conductance on the normalized voltage ${v = eV / \Delta}$.

\begin{figure}[tbp]
\includegraphics[width=1.0\columnwidth]{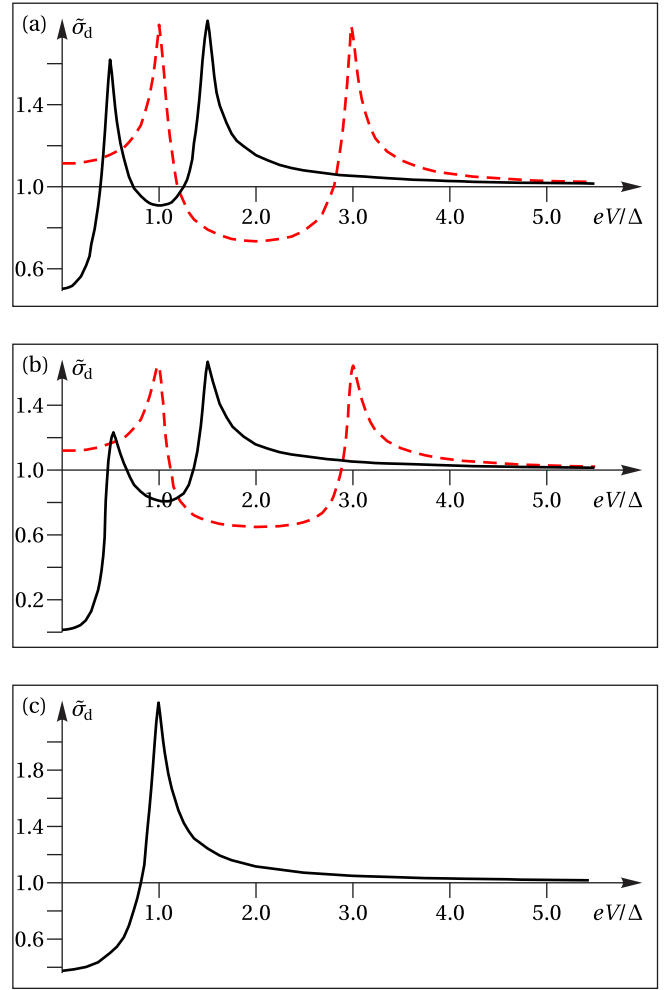}
\caption{(Color online.) The normalized differential conductance at low temperatures (${T \ll \Delta}$) as a function of normalized voltage for the (a)~S$_{\text{m}}$/n/N contact, (b)~S$_{\text{T}}$/n/N (in both cases, the parameters are ${\gamma = 0.3}$, ${h = 0.5}$ for the black solid line and ${h = 5}$ for the red dashed line), and (c)~S/n/N contact, where~S is a BCS superconductor (the parameter is ${\gamma = 0.3}$). Note that the quantities~$\Delta$ and~$h$ are not the true energy gap and the magnetic field, respectively, but the in Eqs.~(\ref{1a})--(\ref{1b}) defined effective values (see also Appendix~\ref{app:Appendix_A}).}
\label{fig:DifCond2a}
\end{figure}

The first term in the denominator, ${\gamma = R_{\text{N}} / R_{\text{S}}}$ determines the resistance of the n/N~interface, while the second term is proportional to the resistance of the interface between the n\nobreakdash-wire and the corresponding superconductor. The first term in the square brackets,~$\nu_{\text{S}}$, determines the conductance of this interface due to quasiparticles with energies above the gap, whereas the second term,~$\gamma f^{2}$, is related to the subgap conductance.

We analyze the differential conductance~$\tilde{\sigma}_{\text{d}}(v)$ and the $I$\nobreakdash-$V$~characteristics~$I(v)$,
\begin{equation}
    I(v) = \int_{0}^{v} \tilde{\sigma}_{\text{d}}(v_{1}) \, \mathrm{d} v_{1} \,, \label{17I_V}
\end{equation}
for contacts of different types. Equations~(\ref{17Con})--(\ref{17I_V}) allow one to calculate the conductance and the $I$\nobreakdash-$V$~characteristics of contacts under consideration. In Fig.~\ref{fig:DifCond2a}, we show the dependence of the normalized differential conductance~$\tilde{\sigma}_{\text{d}}(v)$ on the normalized voltage~$v$ for the three types of contacts, i.e., the S$_{\text{m}}$/n/N contact [Fig.~\ref{fig:DifCond2a}~(a)], the S$_{\text{T}}$/n/N contact [Fig.~\ref{fig:DifCond2a}~(b)], and the S/n/N contact, where~S is a usual BCS superconductor [Fig.~\ref{fig:DifCond2a}~(c)]. Note that the dependence $\tilde{\sigma}_{\text{d}}(v)$ for the case of the BCS superconductor coincides with that for the case of a ``magnetic'' superconductor if one sets ${h = 0}$.

Although the function~$\tilde{\sigma}_{\text{d}}(v)$ in Fig.~\ref{fig:DifCond2a}~(c) looks like the voltage dependence of the differential conductance of an S/I/N junction (where I~stands for an insulating thin layer), it differs from the latter one because this dependence leads to an excess current~$I_{\text{exc}}$. This current is given by the value of~$I_{\text{exc}}(h)$ in Fig~\ref{fig:Exc_Current_on_h}~(a) at ${h = 0}$ (blue dashed line). The appearance of the excess current is a direct consequence of the fact that the integral~$\int_{0}^{\infty} \mathrm{d}v \, [\tilde{\sigma}_{\text{d}}(v)-1]$ is not zero as it takes place in tunnel S/I/N junctions.

It is seen from Fig.~\ref{fig:DifCond2a}~(c) that there is a nonzero subgap conductance in the S/n/N~contact. It is caused by a subgap contribution related to the Andreev reflection. This mechanism is also responsible for a zero-bias peak in the conductance that has been observed in early experiments on S/n/Sm contacts (here, Sm is a n\nobreakdash-doped semiconductor).\cite{Kastalsky91} Theoretical explanations for the observed subgap conductance is given in Refs.~\onlinecite{Volkov92,VOLKOV199321,Hekking_Nazarov_1993}.

In Figs.~\ref{fig:DifCond2a}~(a) and~\ref{fig:DifCond2a}~(b), we plot the voltage dependence of the normalized conductance of the contacts of S$_{\text{m}}$/n/N and S$_{\text{T}}$/n/N~types for different values of~$h$. In both cases, the subgap conductance is not zero, but it is small in contacts of S$_{\text{m}}$/n/N~type if the exchange field~$h$ is small compared to~$\Delta$. The latter property is due to a negligible contribution to the conductance in the subgap region because this contribution is provided by fully polarized triplet Cooper pairs the density of which,~$F_{\text{S}-}$, decreases with decreasing~$h$ since ${F_{\text{S}-} \propto h}$. Note that similar results (nonzero subgap conductance) were obtained in Ref.~\onlinecite{Nagaev99}, where differential conductance of an F'/F/S structure has been studied. However, the case of fully polarized triplet component has not been considered there.

The subgap conductance in another, although similar, systems has been calculated in Ref.~\onlinecite{Beenakker09} on the basis of the scattering matrix approach. The authors considered a half-metal/ferromagnet/superconductor contact in the ballistic regime assuming that the magnetizations in half-metal and ferromagnet are not collinear. They assumed also that only a single conducting channel exists in the system so that the quasiclassical theory can not be applied to the system. To some extent, the results obtained in our paper and in Ref.~\onlinecite{Beenakker09} differ. Although the subgap conductance~$\tilde{\sigma}_{\text{d}}(v)$ calculated in Ref.~\onlinecite{Beenakker09} differs from zero, it turns to zero at ${v = 0}$ whereas~$\tilde{\sigma}_{\text{d}}(0,h)$ obtained by us in the present work is finite.

A similar system consisting of a half-metallic ferromagnet and a superconductor has been studied in Refs.~\onlinecite{Grein_et_al_2010,Lofwander_et_al_2010}. The authors assumed that these materials are separated by a spin-active interface. They also obtained the vanishing zero-bias conductance for ${T \to 0}$. In our case, the finite~$\tilde{\sigma}_{\text{d}}(0,h)$ is caused by unconventional Andreev reflection of triplet Cooper pairs induced in the n\nobreakdash-wire due to proximity effect. This AR make the S$_{\text{m}}$/n~interface partially transparent as it occurs in S/n contacts,\cite{Kastalsky91,Volkov92,VOLKOV199321,Hekking_Nazarov_1993}.

\begin{figure}[tbp]
\includegraphics[width=1.0\columnwidth]{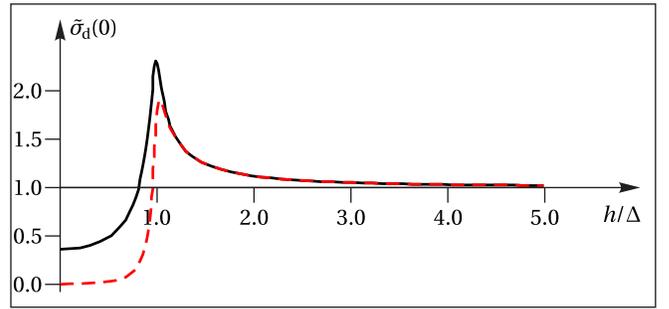}
\caption{(Color online.) Zero-bias conductance as a function of normalized exchange field~$\tilde{h}$ for the contacts for the S$_{\text{T}}$/n/N (black solid line), respectively, S/n/N~contact (red dashed line)---in both cases, the parameter ${\gamma = 0.2}$. Note that the quantities~$\Delta$ and~$h$ are not the true energy gap and the magnetic field, respectively, but the in Eqs.~(\ref{1a})--(\ref{1b}) defined effective values (see also Appendix~\ref{app:Appendix_A}).}
\label{fig:DifCond3}
\end{figure}

The zero-bias conductance~$\tilde{\sigma}_{\text{d}}(0,h)$ as a function of~$h$ is depicted in Fig.~\ref{fig:DifCond3} for the S$_{\text{m}}$/n/N and S$_{\text{T}}$/n/N~contacts. It is equal to zero at ${h = 0}$ in the S$_{\text{T}}$/n/N contact, where only triplet Cooper pairs are present, and has a maximum at ${h = \Delta}$. As mentioned above, at ${h = 0}$ the amplitude of the triplet component turns to zero, and hence the zero-bias conductance vanishes.

\subsection{Excess or deficit current}
\setcounter{paragraph}{0}

We investigate the $I$\nobreakdash-$V$~characteristics of the contacts of the types S$_{\text{m}}$/n/N and S$_{\text{T}}$/n/N.

\paragraph{S$_{\text{m}}$/n/N contact.}

\begin{figure}
  \centering
  \includegraphics[width=1.0\columnwidth]{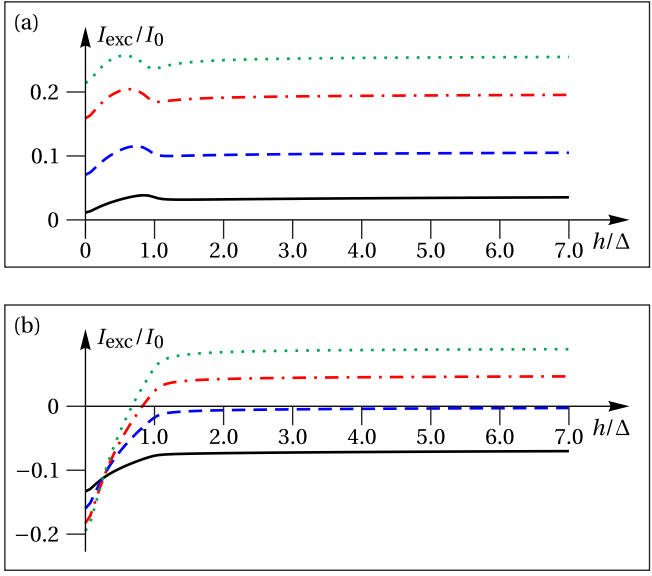}
  \caption{(Color online.) Dependence of the excess, respectively, the deficit current on~$h$ for the (a)~S$_{\text{m}}$/n/N and (b)~S$_{\text{T}}$/n/N contacts. Noticeably is the nonmonotonic behavior of the $I$-$V$~curve in the S$_{\text{m}}$/n/N~contact. The excess current in the S$_{\text{T}}$/n/N contact turns to deficit current at low~${h < h_{\text{c}}}$ (see text). The parameter~$\gamma$ has the values ${\gamma = 0.1}$ (black solid lines), ${\gamma = 0.3}$ (blue dashed lines), ${\gamma = 0.5}$ (red dash-dotted lines), and ${\gamma = 0.7}$ (green dotted lines). The current is normalized to the value of the Ohm's law current at the voltage ${V = 1.0 \Delta/e}$, i.e., ${I_0 = I_{\text{N}}(eV = 1.0\Delta)}$, where ${I_{\text{N}} = V/R}$ with the resistance of the contact in the normal state~$R$. Note that the quantities~$\Delta$ and~$h$ are not the true energy gap and the magnetic field, respectively, but the in Eqs.~(\ref{1a})--(\ref{1b}) defined effective values (see also Appendix~\ref{app:Appendix_A}).}
  \label{fig:Exc_Current_on_h}
\end{figure}

\begin{figure}[tbp]
\includegraphics[width=1.0\columnwidth]{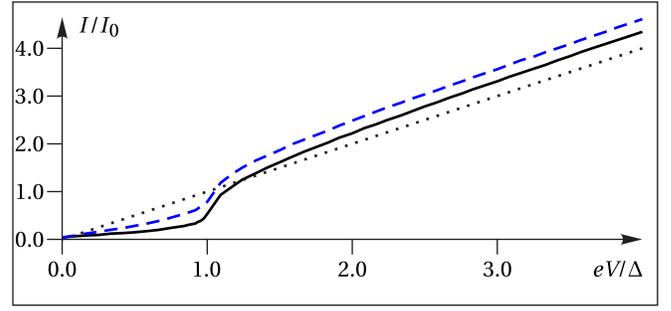}
\caption{(Color online.) Current voltage characteristics for the case of BCS superconductor for ${\gamma = 0.2}$ (black solid line) and ${\gamma = 0.5}$ (blue dashed line). The current is normalized to the value of the Ohm's law current at the voltage ${V = 1.0 \Delta/e}$, i.e., ${I_0 = I_{\text{N}}(eV = 1.0\Delta)}$. The black dotted line indicates the Ohm's law, ${I_{\text{N}} = V/R}$ with the resistance of the contact in the normal state~$R$. Note that the quantities~$\Delta$ and~$h$ are not the true energy gap and the magnetic field, respectively, but the in Eqs.~(\ref{1a})--(\ref{1b}) defined effective values (see also Appendix~\ref{app:Appendix_A}).}
\label{fig:Exc5}
\end{figure}

In the considered case of small but finite~$\gamma$, the $I$\nobreakdash-$V$~characteristics shows an excess current. In particular, for ${h = 0}$ we obtain ${\tilde{I}_{\text{exc}} \propto \gamma \ln (2 / \gamma)}$ [or, with dimension, ${e I_{\text{exc}} (R_{\text{S}} + R_{\text{N}}) \propto \Delta \gamma \ln (2 / \gamma)}$ with ${\gamma = R_{\text{N}} / R_{\text{S}}}$]. The excess current increases with increasing the exchange field~$h$ [see~Fig.~\ref{fig:Exc_Current_on_h}~(a)]. The $I$\nobreakdash-$V$~curve has a simple form for the case ${h = 0}$ (BCS superconductor). For small~$\gamma$ and~${\Gamma \rightarrow 0}$, we obtain
\begin{equation}
    \tilde{I}(v) =  \begin{cases}
                        \gamma \ln \Big( \frac{\sqrt{1 + \gamma^{2}} + v}{\sqrt{1 + \gamma^{2} - v^{2}}} \Big) \,, & v < 1 \,, \\
                        \gamma \ln \Big( \frac{2}{\gamma} \Big) + \sqrt{v^{2} - 1} \,, & v > 1 \,.
                    \end{cases}
                    \label{18I_V}
\end{equation}
In this case, there is an excess current in the $I$\nobreakdash-$V$~curve (see Fig.~\ref{fig:Exc5}).

\paragraph{S$_{\text{T}}$/n/N contact.}

Using Eq.~(\ref{13exc}) we find the excess or deficit current for small~$\gamma$ and~$h$,
\begin{equation}
    \tilde{I}_{\text{exc}} = \frac{\big( \gamma h^{4} \big)^{1/3} c_{3/2}}{2} - \gamma \ln \Big( \frac{2}{e \gamma} \Big) \,,
\label{19exc}
\end{equation}
where ${c_{3/2} = \int_{0}^{\infty} (1 + x^{3/2})^{-1} \mathrm{d} x \approx 1.79}$. One can see that at ${h > h_{c} \equiv \sqrt{\gamma}\big[ \ln (2 / e \gamma) \big]^{3/4}}$, there is an excess current and at ${h < h_{c}}$ the excess current is converted into a deficit current, cf.~Fig.~\ref{fig:Exc_Current_on_h}~(b).

As is seen from Fig.~\ref{fig:Exc_Current_on_h}, the magnitude of the excess current~$I_{\text{exc}}$ in the case of the S$_{\text{T}}$/n/N junction is comparable with the excess current in an S/n/N junction with the same interface resistances. This means that it can be measured experimentally on existing experimental junctions.

\section{Conclusions}

We studied transport properties of ``magnetic'' superconductor~/~normal metal point contacts of different types, in which both the singlet and triplet Cooper pairs are present. It is shown that, as it takes place in point S/n/N~contacts with BCS superconductor, the subgap conductance~$\sigma_{\text{sg}}$ and the excess current~$I_{\text{exc}}$ are not zero even if only fully polarized triplet component exists in the n\nobreakdash-wire. In this case, the~$\sigma_{\text{sg}}$ and~$I_{\text{exc}}$ are caused by an unconventional Andreev reflection without spin flip; the hole moving back along the trajectory of an incident electron with a spin~$\mathbf{S}$ has the same spin direction as~$\mathbf{S}$. A similar AR, equal-spin Andreev reflection, has been studied in a recent paper,\cite{Zhongbo_Wan_2016} where a contact between a ferromagnet and topological superconductor with Majorana modes has been considered.

We considered two types of contacts, namely the S$_{\text{m}}$/n/N~contact, where both the singlet and triplet component exist, and the S$_{\text{T}}$/n/N~contact, in which only fully polarized triplet Cooper pairs penetrate into the n\nobreakdash-wire. In both types of contacts, the subgap conductance and the excess current are present. In the second type of contacts, in S$_{\text{T}}$/n/N, these are caused by an equal-spin AR. With decreasing the magnitude of the exchange field~$h$ the excess current in the S$_{\text{T}}$/n/N~contact is transformed into a deficit current~$I_{\text{def}}$. The systems considered by us can be realized experimentally taking into account a rapid progress in preparing S/F nanostructures of different kinds.\cite{Singh_et_al_2016_arXiv,Birge12,Halasz_et_al_2011} The obtained results can be used for identifying the long-range triplet component and in future applications in spintronics.\cite{Linder_Robinson_2015}

\acknowledgments

We appreciate the financial support from the DFG via the Projekt~EF~11/8\nobreakdash-2; K.~B.~E.~gratefully acknowledges the financial support of the Ministry of Education and Science of the Russian Federation in the framework of Increase Competitiveness Program of  NUST~``MISiS'' (Nr.~K2-2014-015).

\appendix

\section{Green's function in an S/F bilayer}
\label{app:Appendix_A}

We consider an F/S~bilayer and show that, under certain conditions, the matrix Green's function~$\hat{g}_{\omega}$ coincides with that in a superconductor with a built-in exchange field~$h$. We assume that the thickness of the F~layer~$d_{\text{F}}$ is small so that the condition, ${d_{\text{F}} \ll \xi_{h} \equiv \sqrt{D/h}}$, is fulfilled. Then, the Usadel-like equation~(\ref{2}) in the F~region can be integrated over the thickness and we come to Eq.~(\ref{15Usadel}) with ${\hat{\Lambda}_{\pm} = {\hat{\tau}_{3}} [G_{\text{S}} + \omega/\epsilon_{\text{sg}} \pm ih / \epsilon_{\text{sg}}] + {\hat{\tau}_{2}} F_{\text{S}}}$, where ${G_{\text{S}} = F_{\text{S}} = \omega/\zeta_{\omega}}$ is the Green's function in~S, ${\zeta_{\omega} = \sqrt{\omega^{2} + \Delta^{2}}}$. The exchange field vector~$\mathbf{h}$ is set along the $z$~axis. The subgap energy~$\epsilon_{\text{sg}}$ is defined in Eq.~(\ref{1b}) and we use the Matsubara representation. The matrix~$\hat{g}_{\omega}$ is diagonal in the spin-space with elements~$\hat{g}_{\omega \pm}$. The retarded Green's function~$\hat{g}^{R}$ can be directly obtained from $\hat{g}_{\omega}$ using the relation ${\omega = -i (\epsilon + i 0)}$. The solution for~$\hat{g}_{\omega}$ can be easily found as in Sec.~\ref{sec:2a} and has the form
\begin{equation}
\hat{g}_{\omega \pm} = \frac{\hat{\Lambda}_{\pm}}{a_{\pm}} \,, \label{A1}
\end{equation}
where ${a_{\pm} = \sqrt{(G_{\text{S}} + \omega/\epsilon_{\text{sg}} \pm ih/\epsilon_{\text{sg}})^{2} + F_{\text{S}}^{2}}}$. If the resistance of the F/S~interface is large enough (this corresponds to real experiments) so that the subgap energy ${\epsilon_{\text{sg}}}$ is small in comparison with~$\Delta$, the solution Eq.~(\ref{A1}) can be written as
\begin{equation}
\hat{g}_{\omega \pm} = \frac{{\hat{\tau}_{3}} [\omega \pm ih] + {\hat{\tau}_{2}} \epsilon_{\text{sg}}}{\sqrt{(\omega \pm ih)^{2} + \epsilon_{\text{sg}}^{2}}} \,.
\label{A2}
\end{equation}
Equation~(\ref{A2}) shows that the Green's function in the F~layer has the same form as in a superconductor with the energy gap~$\epsilon_{\text{sg}}$ and built-in exchange field~$h$. This equation is valid if the thickness~$d_{\text{F}}$ satisfies the condition
\begin{equation}
\frac{D \kappa_{\text{SF}}}{\Delta} \ll d_{\text{F}} \ll \sqrt{\frac{D}{h}} \,. \label{A3}
\end{equation}
As follows from this condition, the exchange field~$h$ can be much larger than the effective energy gap~$\epsilon_{\text{sg}}$.

%

\end{document}